\documentclass[pra,twocolumn,showpacs,preprintnumbers,amsmath,amssymb]{revtex4}
\usepackage{graphicx}
\usepackage{dcolumn}
\usepackage{bm}
\usepackage{latexsym}
\usepackage{amssymb}
\usepackage{amsfonts}
\usepackage{amsmath}

\begin{document}

\title{Component separation of two-component fermion clouds in a spin-dependent external potential
by spin-density-functional theory}
\author{Gao Xianlong}
\email{gaoxl@zjnu.edu.cn}
\affiliation{Department of Physics, Zhejiang Normal University, Jinhua 321004, China}
\date{\today}
\begin{abstract}
We investigate the component separation in one-dimensional two-component fermion clouds in a spin-dependent external potential. The density distributions and the state diagram are studied by means of spin-dependent density-functional theory. The component separation between spin-up and spin-down atoms is induced by the interplay of the spin-dependent harmonic confinement and the strong repulsive interaction between the inter-components. We find the existence of a threshold repulsive interaction strength above which the component separation evolves. Different state diagrams are mapped out numerically, from which two regions are distinguished, i.e., the phase-mixed region with both spin-up and spin-down mixtures in the center of the trap and the phase-separated region with only spin-up atoms remaining in the center.
\hspace*{7.0cm}
\end{abstract}


\maketitle
%%%%%%%%%%%%%%%%%%%%%%%%%%%%%%%%%%%%%%%%%%%%%%%%%%%%%%%%%%%%%%%%%%%%%%
\section{Introduction}
Much progress has been achieved recently in the field of ultracold atomic gases~\cite{review paper}, among which the ground state of one-dimensional (1D) systems of fermionic atoms trapped in a harmonic trap has been the subject of numerous analytical and numerical studies~\cite{1DFermigasestheory,BALDA} since fermion gases trapped in `atomic quantum wires' is realized experimentally~\cite{moritz_prl_2005} and cooled to temperatures $T \sim 0.1~T_{\rm F}$, where $T_{\rm F}$ is the Fermi temperature~\cite{liao_naturephys_2010}.

A strategy currently used for cooling fermion atomic gases is sympathetic cooling between the fermions and a second gaseous component made either of fermions in a different internal state or of bosons via s-wave collisions~\cite{Das,trap-imbalance1}. At increasing values of the scattering lengths, the boson-fermion cloud may undergo demixing~\cite{Das,FBM_demix1,Cazalilla,Imambekov,Yin,Wang,BM_demix2}. When the trapping potentials become component-dependent~\cite{BM_demix2,Dalmonte}, due to the different masses or the magnetic oscillator frequencies~\cite{trap-imbalance2,Jin}, the increasing repulsive interaction between the components will also demix them~\cite{Karpiuk,FM_demix1,pra-spin}. Locating the onset of incipient spatial separation, i.e., the component separation point, is relevant to fermion cooling, since at that point the diminished overlap between the two clouds will reduce the effectiveness of the collisional transfer processes.

A great deal of research has touched the topic of 1D atomic mixtures in optical lattices, modeled by a lattice Hamiltonian with confining potentials of atoms interacting through a Hubbard-type term~\cite{BM_demix2,pra-spin,Dalmonte,FM_demix1}, or mixtures in a continuum space, modeled by Gaudin-Yang Hamiltonian interacting through a contact short-range term~\cite{trap-imbalance2,GYmodel}. When detuning asymmetrically the laser frequencies with respect to the two hyperfine states~\cite{trap-imbalance1}, or when trapping the two-component atomic gases of unequal masses, one needs to consider the confined external potentials to be spin-dependent~\cite{Dalmonte,trap-imbalance2,pra-spin}.

Many theories have been tried in understanding the rich quantum phases in the atomic mixtures. The mean-field theory (reliable at weak interaction) and the Luttinger liquid (valid asymptotically at small momenta and low energies) have predicted the occurrence of component separation, i.e., demixing of the two components in spatial space~\cite{Cazalilla,Das,Meanfield}. State diagrams are computed for two-component one-dimensional quantum gases using density-matrix renormalization group techniques~\cite{Roux}. The local density approximation based on the Bethe-ansatz solution shows that the harmonically trapped 1D mixtures partially demix at strong repulsive interaction~\cite{Imambekov,Yin}. A density-functional theory for the 1D harmonically trapped Bose-Fermi mixture with repulsive contact interactions is recently used to study the component separation~\cite{Wang}. For two-component fermionic mixtures of same masses, a Bethe-ansatz based spin-density-functional theory (SDFT) has been successfully used in studying the static and dynamic properties~\cite{BALDA}, which is suitable for the whole interaction range without limitations from the particle number and system size~\cite{gaosubmitted,WangJJ}. This method has never been used in studying the component separation induced by the spin-dependent external potential, which is the purpose of this paper.

For a better understanding of the effects of the spin-dependent external potentials, the repulsive interaction, and the polarization on the process of demixing at a large range of parameters, it is essential to have a complete state diagram, from which one can easily find the optimal parameters to realize the demixing or to control the cooling efficiency. In this paper, we study the demixing of the two fermion species (taking as pseudospins) of same masses in a continuum space in the presence of spin-dependent external potentials.

The contents of the paper are arranged as follows. In Sect.~\ref{sect:model} we introduce the model: an inhomogeneous Gaudin-Yang Hamiltonian of a contact interaction. Then we briefly summarize the self-consistent spin-density-functional scheme used to deal with the inhomogeneous system. In Sect.~\ref{sect:numerical_results} we report and discuss our main numerical results. At last, a concluding section summarizes our results.

\section{Inhomogeneous Gaudin-Yang model and spin-density-functional theory}
\label{sect:model}
We consider a two-component Fermi gas with $N_f$ atoms of same mass $m$ confined inside a strongly elongated harmonic trap along the $x$-direction.
The two species of fermionic atoms are assumed to have different pseudospin $\sigma=\uparrow$ or $\downarrow$ (hyperfine-state label). The number of atoms of spin $\sigma$ is $N_\sigma$ satisfying $N_\uparrow+N_\downarrow=N_f$. The trapping potential is axially symmetric and characterized by angular frequencies $\omega_\perp$ and $\omega_\sigma$ in the radial and longitudinal directions, respectively, with $\omega_{\downarrow},\,\omega_{\uparrow} \ll \omega_\perp$.

The gas is dynamically 1D if the anisotropy parameter of the trap is much smaller than the inverse atom number, $\omega_\uparrow/\omega_\perp,\,\omega_\downarrow/\omega_\perp\ll N^{-1}_f$. It can thus be described by an inhomogeneous Gaudin-Yang Hamiltonian,
\begin{eqnarray}\label{eq:igy}
{\hat {\cal H}} &=& {\hat {\cal T}} + {\hat {\cal V}} + {\hat {\cal W}} = -\frac{\hbar^2}{2m}\sum_\sigma\int_{-\infty}^{+\infty}dx~{\hat \Psi}^\dagger_\sigma(x) \partial^2_x {\hat \Psi}_\sigma(x) \nonumber\\
&+& g_{\rm 1D}\int_{-\infty}^{+\infty}dx~{\hat \Psi}^\dagger_{\uparrow}(x){\hat \Psi}^\dagger_{\downarrow}(x){\hat \Psi}_{\downarrow}(x){\hat \Psi}_{\uparrow}(x)  \nonumber\\
&+& \frac{1}{2} m\sum_{\sigma}\omega^2_\sigma \int_{-\infty}^{+\infty}dx~{\hat \Psi}^\dagger_\sigma(x) x^2 {\hat \Psi}_\sigma(x)~,
\end{eqnarray}
where ${\hat \Psi}^\dagger_\sigma(x)$ [${\hat \Psi}_\sigma(x)$] is a field operator that creates (annihilates) a fermion of spin $\sigma$ at position $x$ and $g_{\rm 1D} \simeq 4\hbar^2 a_{\rm sc}/(m a^2_\perp)$ (in the limit $a_{\rm sc} \ll a_\perp$) is a parameter that determines the strength of inter-particle repulsions~\cite{olshanii_prl_1998}. Here $a_\perp=(\hbar/m\omega_\perp)^{1/2}$ is the harmonic-oscillator length. The 3D scattering length $a_{\rm sc}$ can be tuned easily using a magnetic field~\cite{moritz_prl_2005}. The first term in Eq.~(\ref{eq:igy}) (${\hat {\cal T}}$) is the kinetic energy whereas the second term (${\hat {\cal V}}$) describes two-body short-range interactions between spin-up and spin-down atoms. Finally, the third term (${\hat {\cal W}}$) is a spin-dependent parabolic trapping potential.

We define the dimensionless ratio of the spin-up and spin-down external potential as,
\begin{equation}\label{eq_gamma}
\gamma=\frac{\omega^2_{\downarrow}}{\omega^2_{\uparrow}}~.
\end{equation}
In our study, we assume $\omega_\downarrow<\omega_\uparrow$ by fixing $\omega_{\uparrow}=1$ and thus $0<\gamma\le 1$.

We choose the unit of length as the oscillator length for the spin up atoms $\ell_{\rm \uparrow}=(\hbar/m\omega_\uparrow)^{1/2}$, and $\hbar\omega_\uparrow$ as unit of energy, the Hamiltonian ({\ref{eq:igy}) is governed by
the dimensionless coupling parameter
\begin{equation}\label{eq_lambda}
\lambda=\frac{g_{\rm 1D}}{\ell_{\rm \uparrow}\hbar\omega_\uparrow}~.
\end{equation}
In this work we focus our attention on the interplay of $\lambda$ and $\gamma$ for inter-atom repulsions ($\lambda >0$) on the local spin-resolved density, $n_{\sigma}(x) = \langle
{\hat \Psi}^\dagger_\sigma(x){\hat \Psi}_\sigma(x)\rangle$,
the total density $n(x) = \sum_\sigma n_\sigma(x)$, and the local magnetization $\zeta(x) = [n_\uparrow(x)-n_\downarrow(x)]/2$.

For $\omega_{\sigma}=0$, the Hamiltonian (\ref{eq:igy}) returns to the homogeneous Gaudin-Yang model analytically solvable by means of the Bethe-ansatz technique~\cite{Burke,Gaudin,BALDA} and determined by
the linear density $n=N_f/L$, by the spin polarization $\zeta=(N_{\uparrow}-N_{\downarrow})/N_f$, and by the interaction strength $g_{\rm 1D}$.

\begin{figure*}
\begin{center}
\tabcolsep=0 cm
\begin{tabular}{cc}
\scalebox{0.35}[0.35]{\includegraphics{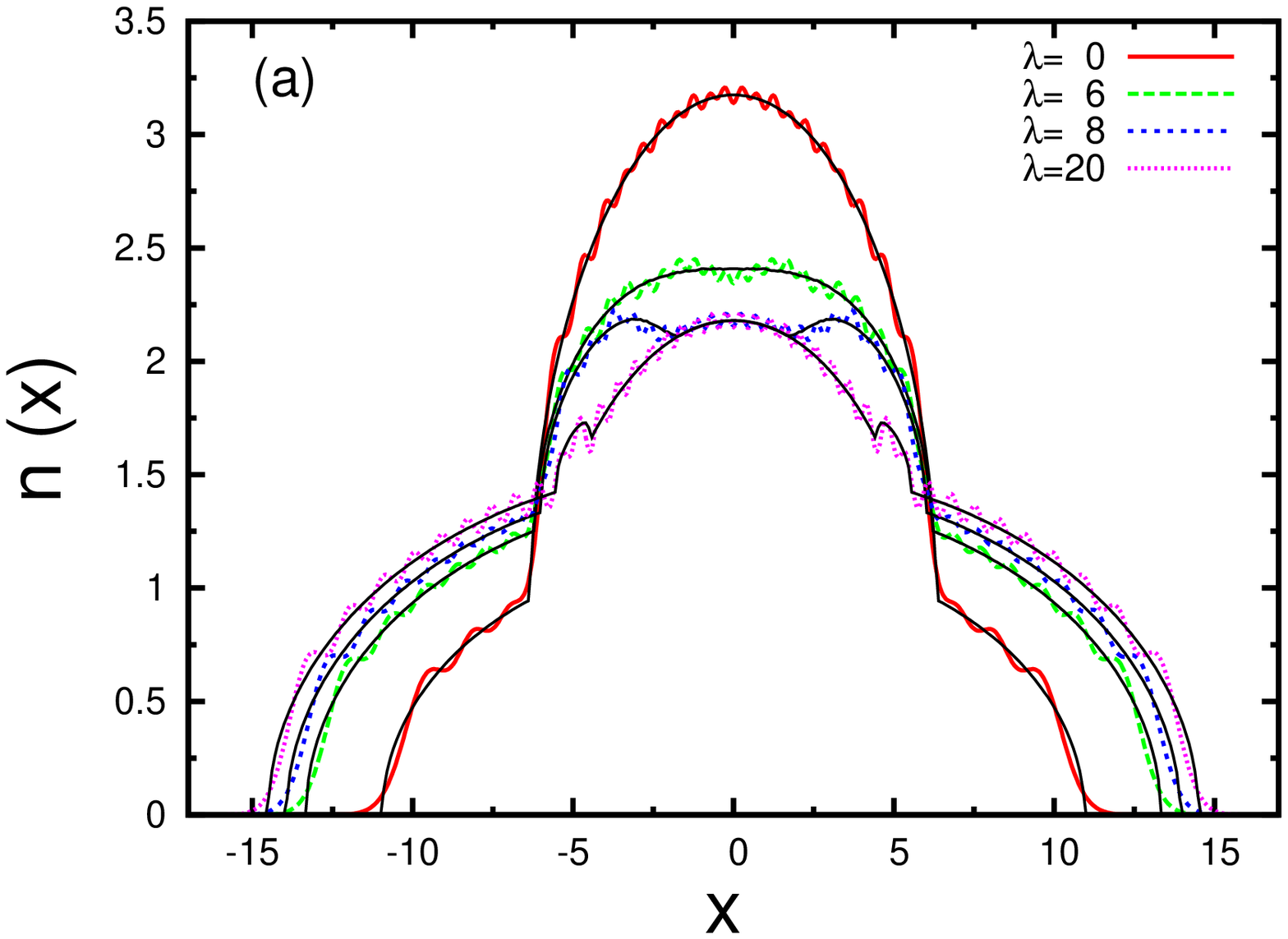}}&
\scalebox{0.35}[0.35]{\includegraphics{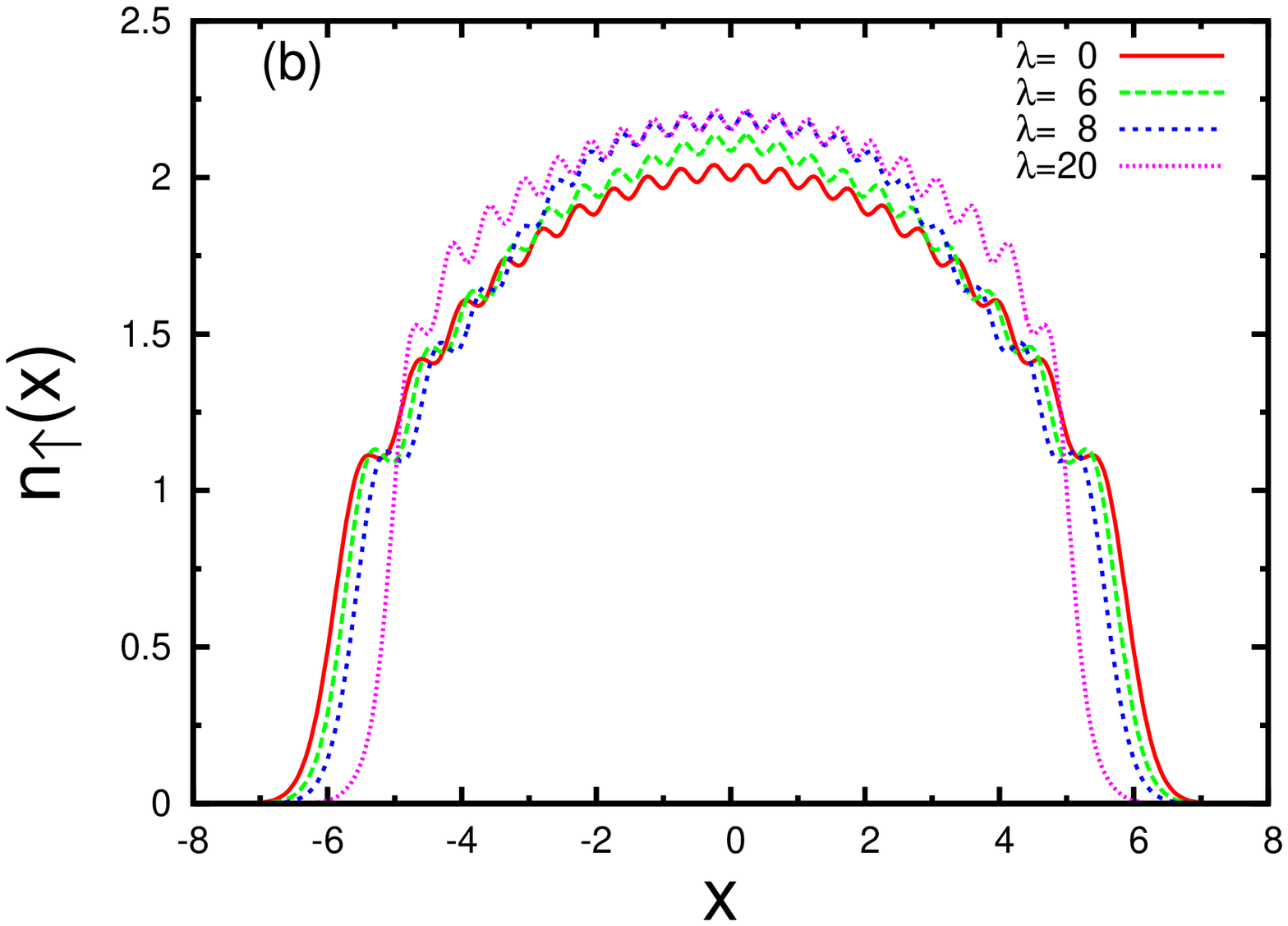}}\\
\scalebox{0.35}[0.35]{\includegraphics{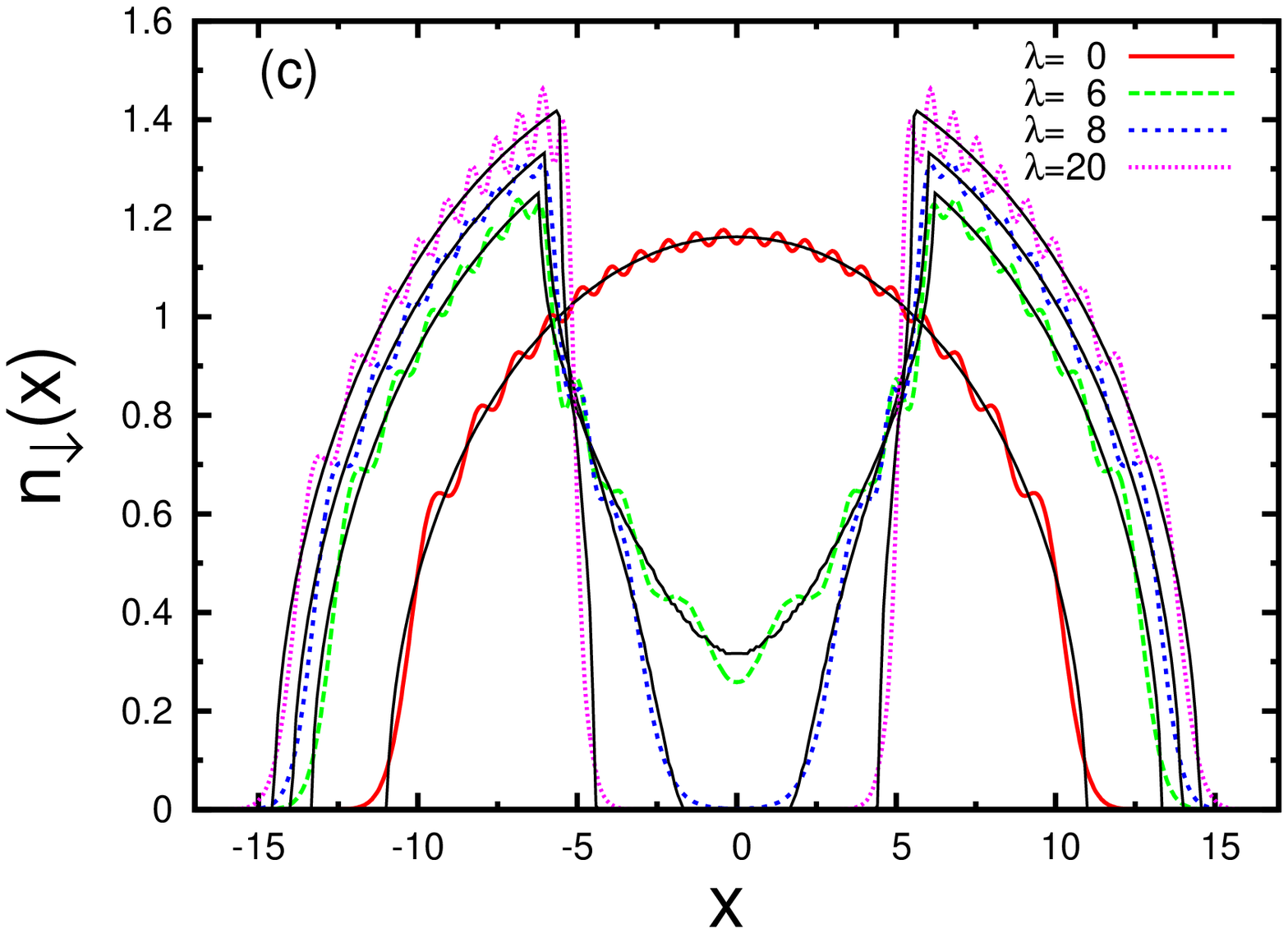}}&
\scalebox{0.35}[0.35]{\includegraphics{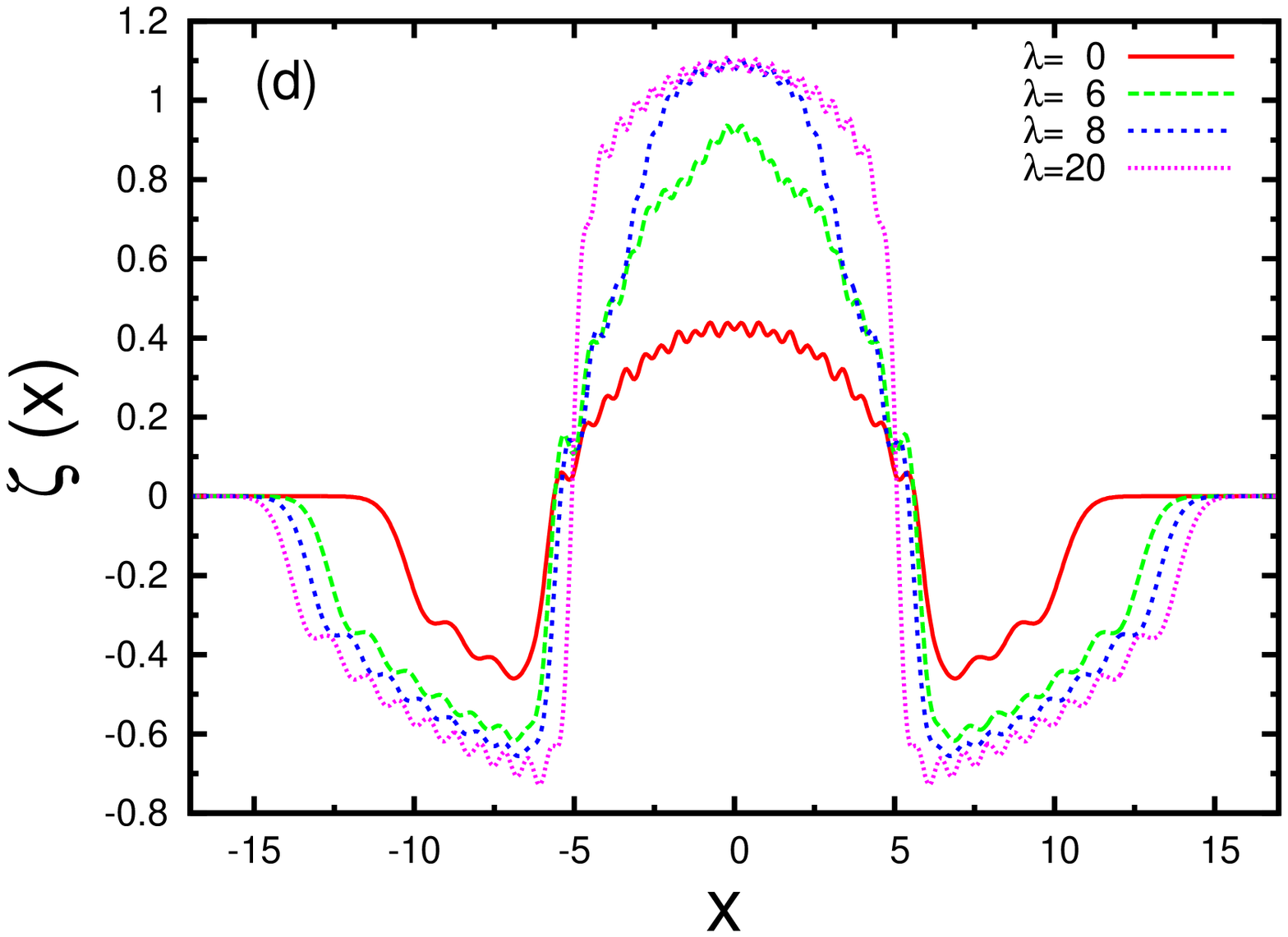}}
\end{tabular}
\caption{(Color online) (a) The total ground-state density distributions $n(x)$; (b) spin-resolved densities for spin-up $n_\uparrow(x)$; (c) spin down $n_\downarrow(x)$ atoms; and (d) the local magnetization (in units of $\ell^{-1}_\uparrow$) as a function of $x$ (in units of $\ell_{\uparrow}$) for a unpolarized Fermi gas with $N_\uparrow=N_\downarrow=20$ and $\gamma=1/9$.
For comparison, the non-interaction case ($\lambda=0$) is also shown in the figure.
In (a) and (c), the mean-field results based on the parametrized energy functional from G. Xianlong and R. Asgari, \pra {\bf 77}, 033604 (2008) are included by thin black lines.
\label{fig:one}}
\end{center}
\end{figure*}
\begin{figure}
\begin{center}
\includegraphics*[width=1.00\linewidth]{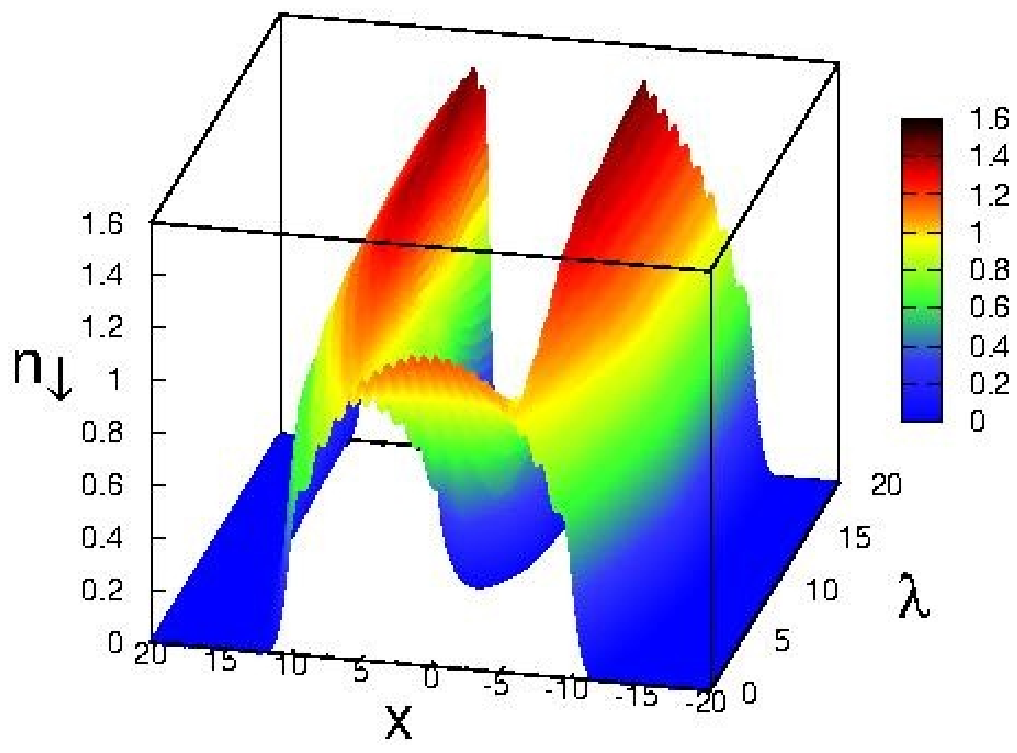}
\caption{(Color online) The 3D plot of the spin-down densities $n_\downarrow(x)$ (in units of $\ell^{-1}_\uparrow$) as functions of $x$ (in units of $\ell_{\uparrow}$) and $\lambda$. The parameters are the same as those in Fig. \ref{fig:one} but with continuously varying interaction strength in the range [0, 20].
\label{fig:two}}
\end{center}
\end{figure}

For $\gamma=1$, the Hamiltonian (\ref{eq:igy}) is equivalent to the inhomogeneous Gaudin-Yang model in a spin-independent external potential, which is extensively studied in Refs.~\cite{BALDA},~\cite{GYmodel}, and~\cite{gaosubmitted}. In this case, the system undergoes the crossover from the $2k_F$-Friedel to $4k_F$-Wigner oscillations (with $k_{\rm F}$ the Fermi wave vector) at a strong interaction strength. However, there is no component separation, however strong the repulsive interaction is~\cite{gaosubmitted}. The spin-up and spin-down densities are always locally the same. That is, $\zeta(x)\equiv 0$ or $n_\uparrow(x)=n_\downarrow(x)$. At $\zeta=1$, the system is composed of fully spin-polarized Fermi gases in a normal phase.

Originally, static density-functional theory (DFT) is formulated for many-electron systems in the continuous space of long-range Coulomb interaction. On the other hand, DFT can be also used in the model system of contact interaction~\cite{Gunnarsson,Lima,Schenk,Burke,Vieria,BALDA}. In the DFT language, the ground-state spin-density distributions $n_\sigma(x)$ can be calculated by solving self-consistently the Kohn-Sham (KS) equation,
\begin{equation}\label{eq:kss}
\left[-\frac{\hbar^2}{2m}\frac{\partial^2}{\partial x^2}+V^{(\sigma)}_{\rm KS}[n_\sigma](x)\right]\varphi_{\alpha,\sigma}(x)=\varepsilon_{\alpha,\sigma}\varphi_{\alpha,\sigma}(x)
\end{equation}
with the KS orbital $\varphi_{\alpha,\sigma}$ satisfying
\begin{equation}\label{eq:closure}
n_\sigma(x)=\sum_{\alpha=1}^{N_\sigma}\left|\varphi_{\alpha,\sigma}(x)\right|^2\,.
\end{equation}
Here, $V^{(\sigma)}_{\rm KS}[n_\sigma](x)=V^{(\sigma)}_{\rm H}[n_\sigma](x)+V^{(\sigma)}_{\rm xc}[n_\sigma](x)+V^{(\sigma)}_{\rm ext}(x)$ is the spin-dependent effective KS potential, where $V^{(\sigma)}_{\rm H}$ is the mean-field term $V^{(\sigma)}_{\rm H}=g_{\rm 1D}n_{\bar \sigma}(x)$ with ${\bar \sigma}$ the opposite spin of $\sigma$, $V^{(\sigma)}_{\rm xc}[n_\sigma](x)$ is the exchange-correlation potential defined as the functional derivative of the exchange-correlation energy $E_{\rm xc}[n_\sigma]$ evaluated at the ground-state density profile, $V^{(\sigma)}_{\rm xc}(x)=\delta E_{\rm xc}[n_\sigma]/\delta n_\sigma(x)|_{\rm \scriptscriptstyle GS}$. $V^{(\sigma)}_{\rm ext}(x)=m\omega_\sigma x^2/2$ is the spin-dependent external potentials. In this paper, we work in the canonical ensemble by keeping the total number of atoms constant and varying the number of spin-up and spin-down atoms in the system, $N_\sigma=\int dx n_\sigma(x)$.

To solve the KS equation~(\ref{eq:kss}) together with~(\ref{eq:closure}), the only term one needs to approximate is the exchange-correlation functions $E_{\rm xc}[n_\sigma]$, which is normally done by taking the local-spin-density approximation (LSDA). In the following we employ a Bethe-ansatz-based LSDA (BALSDA) functional for the exchange-correlation potential,
\begin{eqnarray}\label{eq:balda}
E_{\rm xc}[n_\sigma] & \approx & E^{\rm LSDA}_{\rm xc}[n_\sigma]\\
&=&\int dx\, n(x)\left.\varepsilon^{\rm hom}_{\rm xc}(n,\zeta,g_{\rm 1D})\right|_{n\rightarrow n(x),\zeta\rightarrow \zeta(x)}\,,\nonumber
\end{eqnarray}
where the exchange-correlation energy per particle $\varepsilon^{\rm hom}_{\rm xc}$ of the homogeneous Gaudin-Yang model is defined by
\begin{eqnarray}\label{eq:xc}
\varepsilon^{\rm hom}_{\rm xc}(n,\zeta,g_{\rm 1D})&=&\varepsilon_{\rm GS}(n,\zeta,g_{\rm 1D})-\kappa(n,\zeta)\nonumber\\
&-&\varepsilon_{\rm H}(n,\zeta,g_{\rm 1D})\,.
\end{eqnarray}
Here $\varepsilon_{\rm GS}(n,\zeta,g_{\rm 1D})$ is the ground-state energy of the exact Bethe-ansatz solution of the model,
$
\varepsilon_{\rm H}(n,\zeta,g_{\rm 1D})=\frac{1}{4}g_{\rm 1D}n^2(1-\zeta^2)
$
is the mean-field energy, and
$\kappa(n,\zeta)=\pi^2\hbar^2 n^2(1+3\zeta^2)/24m$ is the noninteracting kinetic energy~\cite{Gaudin}.
LDA/LSDA is known to provide an excellent description of the ground-state properties of a variety of inhomogeneous systems~\cite{Giuliani_and_Vignale}. The Bethe-ansatz based LDA/LSDA has been successfully used in calculating the static and dynamic properties of the systems of contact interactions\cite{Burke,BALDA}.

\section{Numerical results}
\label{sect:numerical_results}
In the following we study the effects of $\lambda$ and $\gamma$ on the component separation by fixing the total number
of fermions $N_f=40$. The $\zeta$ is adjusted by varying the corresponding $N_\uparrow$ and $N_\downarrow$. Due to the finite-size nature of the system studied in this paper, the $\zeta$ is restricted to be $\zeta \in [0, 0.9]$.

Firstly, we study the component separation induced by the repulsive interaction $\lambda$ for fixed ratio of the spin-dependent parabolic potentials $\gamma=1/9$. In Fig.~\ref{fig:one} (a)-(d), the total atomic density, the spin-resolved density for spin-down and spin-up atoms, and the local magnetization are shown, respectively.
We illustrate the effects of the repulsive interaction on the local density distributions and the local magnetization.
For the total density, the effect of the repulsive interaction makes it lower and broader, as expected.
At weak interactions, both spin-up and spin-down atoms are located in the center of the trap. It is a phase-mixed (PM) region.
With the increasing repulsive interactions, the density of the spin-up atoms in a tighter confining potential, becomes shallower and higher, however, growing at a slow pace with increasing interaction energy. As a result, spin-down atoms are excluded from the center of the trap, but in a dramatic way, to decrease the interaction energy while the potential energy is increased. The equilibrium density profiles are the result of the competition between these two opposite effects. We show that, there exists a threshold beyond which the total energy is minimized by a configuration in which the two components are spatially separated. Accordingly we define the threshold where spin-down atoms are depleted completely from the center as the phase-separated (PS) by requiring $n_\downarrow(0)\lesssim 10^{-3}$. Considering that the density in the trap center may oscillate,
we can also define the PS region determined by $\int^{\Delta x}_{-\Delta x} dx n_\downarrow(x)\lesssim 10^{-3}$
with $\Delta x=0.1$. We have checked that in this case the phase boundary does not change qualitatively.
In the present case, the onset of the PS region happens at a threshold interaction strength $\lambda_c=8.17$.

With increasing $\lambda$, the local magnetization $\zeta(x)$ in the central region, becomes stronger and stronger. When a PS region is achieved, the fermion clouds in the trap center are composed of the fully polarized spin-up fermions.
For $\gamma \ne 0$, a flat region of $\zeta (x)$ is seen in the center of the trap and two dips are shown at the edges with the excess spin-down atoms. The increase of the repulsive interaction strength shows a signature that $\zeta (x)$ is more negative at the edges, that is, more and more spin-down atoms are repelled from the center of the trap and accumulate at the periphery region.
For a strong repulsive interaction where the component separation begins to evolve, $\zeta (x)$ changes from negative to positive with a large slope.

In Fig.~\ref{fig:one} (a) and Fig.~\ref{fig:one} (c), we also include the mean-field results. Comparing to the BALSDA scheme, not only the exchange-correlation energy, but also the noninteracting kinetic energy are treated locally~\cite{BALDA}. We find that the mean field gives qualitatively the same results as those of BALSDA. However, the performance of the mean-field scheme at weaker interactions deteriorates with decreasing particle number where the kinetic energy processes. As a result, the regions close to the edges of the trap becomes less accurate. In the system of spin-dependent external potential, the phase-separation areas where the densities become small are also those where the mean field is less accurate.

To have a clear demonstration on how spin-down atoms are repelled from the center of the trap while increasing the repulsive interaction, in Fig. \ref{fig:two}, the 3D plot of the spin-down densities $n_\downarrow(x)$ is shown as functions of position $x$ and interaction strength $\lambda$. With the increasing of the repulsive interactions $\lambda$, spin-down atoms are depleted gradually from the center of the trap. Further increasing the interactions, the two components fully separate, i.e., occupy different regions of space. The respective density shapes of the spin-up and spin-down atoms become stable at much stronger interaction strength. In this example, we find the densities for spin-up and spin-down atoms remain the same at $\lambda \gtrsim 20$.

\begin{figure}
\begin{center}
\includegraphics[width=1.00\linewidth]{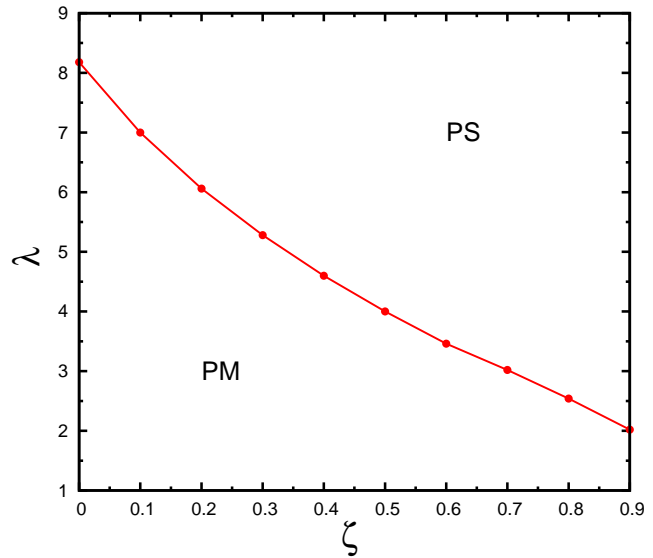}
\caption{(Color online) The $(\lambda, \zeta)$ state diagram at a fixed ratio $\gamma=1/9$. As the polarization increases, the fermionic systems under spin-dependent external potentials undergoes a crossover from a PM to a PS region. The line serves as a guide for the eyes.
\label{fig:three}}
\end{center}
\end{figure}
\begin{figure}
\begin{center}
\includegraphics[width=1.00\linewidth]{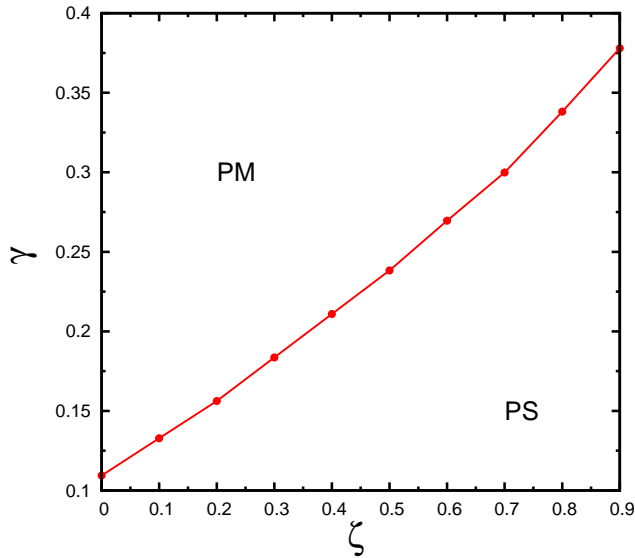}
\caption{(Color online) The $(\gamma, \zeta)$ state diagram at a fixed strong repulsive interaction of $\lambda=8$. As the polarization increases, the fermionic systems in the presence of spin-dependent external potentials undergoes a crossover from a PM to a PS region. The line serves as a guide for the eyes.
\label{fig:four}}
\end{center}
\end{figure}
\begin{figure}
\begin{center}
\includegraphics[width=1.00\linewidth]{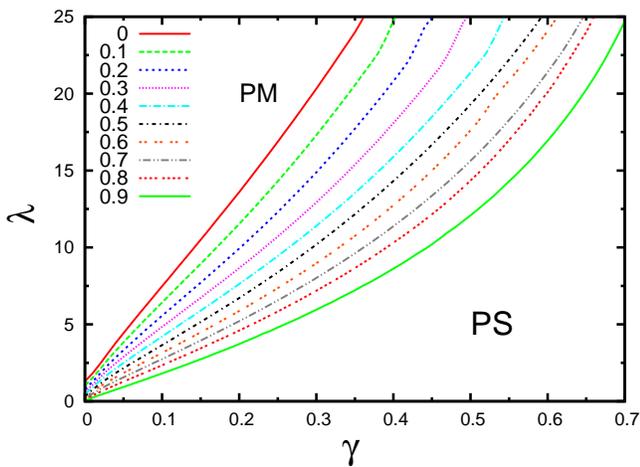}
\caption{(Color online) The $(\lambda, \gamma)$ state diagram with varying polarizations $\zeta$ from $\zeta=0$ to $\zeta=0.7$.
\label{fig:five}}
\end{center}
\end{figure}

To understand the influence of the polarization on the demixing process, we numerically map out the different state diagrams in Figs. \ref{fig:three}-\ref{fig:five}. Two regions are seen: the PM region with spin-up and spin-down mixtures in the center of the trap and the PS region with only spin-up atoms left in the center. The crossover to the PS regime is smooth in the present finite size system of confined gases. The transition between these two regimes as functions of the physical system parameters will be explained as follows.

The state diagram as a function of $\lambda$ and $\zeta$ at a fixed ratio $\gamma=1/9$ is shown in Fig. \ref{fig:three}.
At a certain polarization $\zeta$, the system is in a PM region at weak interaction and in a PS region at strong interaction. When the polarization becomes larger, the demixing is easier due to more spin-up atoms and less spin-down atoms in the trap, and, consequently, a smaller threshold value for the interaction strength $\lambda_c$ is needed for component separation. As a result, the phase boundary in the $(\lambda, \zeta)$ state diagram is a monotonically decreasing curve.

Now, let us concentrate on the component separation induced by spin-dependent parabolic potentials and the polarization at fixed strong interaction of $\lambda=8$, which is described in the $(\gamma, \zeta)$ state diagram in Fig. \ref{fig:four}.
At a certain polarization $\zeta$, the system is in a PM region at a larger ratio $\gamma$ and in a PS region at a smaller $\gamma$ (i.e., the bigger difference between the oscillator frequencies $\omega_\uparrow$ and $\omega_\downarrow$). Similarly, the polarization makes the PS state easier, which explains why the phase boundary monotonically increases.
The phase boundaries in Figs.~\ref{fig:three} and ~\ref{fig:four} can be extended to $\zeta\rightarrow 1$. However, at $\zeta=1$
the system is a trivial fully spin-polarized Fermi gas in a normal phase.

In Fig. \ref{fig:five}, at varying polarizations, the $(\lambda, \gamma)$ state diagram is shown. The region above the cures gives a PS state, while the one below is a PM state. For a fixed polarization, the smaller the $\gamma$, i.e., the tighter the confinement for spin-up atom, the easier is to deplete the spin-down atoms. As a result, smaller $\lambda_c$ is needed to achieve PS. This is the reason why the phase boundary is an increasing function of the ratio $\gamma$. Compared to the different polarization, we find that, the bigger polarization, the easier to deplete the spin-down atoms, consistent with what is described in Fig.~~\ref{fig:three}. Thus, a larger PS region is obtained.

\section{Conclusions}
In summary, by adopting a Bethe-ansatz based spin-density-functional method, in this paper we have performed a detailed numerical study of a 1D Gaudin-Yang model in a spin-dependent harmonic trap. The interplay among the repulsive interaction, the spin-dependent harmonic traps, and the polarization is studied. We find that, for the system in the spin-dependent external potentials, there exists a threshold value for the interaction strength beyond which a component separation occurs with two Fermi components staying in the different spatial regions. For the system with a weak interaction strength, upon increasing the trap imbalance, the spin-up atoms are confined more and more in the center of the trap and a depletion occurs for the spin-down atoms due to the increasing interaction energy. With a fixed ratio of the external potentials $\gamma$, when the interaction strength $\lambda$ is larger than a threshold value $\lambda_c$, the competition of the interaction energy and potential energy leads to a phase-demixed region.

For a polarized system but fixed total atoms, we obtained a $(\lambda, \zeta)$ state diagram at a fixed $\gamma$, a $(\gamma, \zeta)$ state diagram at a fixed $\lambda$, and a $(\lambda, \gamma)$ state diagram with varying $\zeta$, from which it is easy to judge in which parameters the system is in a PS region.

The state diagrams provide the actual range of parameters about the onset of incipient spatial separation and help us find an optimal parameters to demix the two components. In the process of the sympathetic cooling, we can make use it to control the cooling efficiency since the collision rate is strongly related to the overlapping region between the two components.

Experimentally, when selectively trapping atoms of the same species in different hyperfine levels, such as for $^{6}$Li-$^{6}$Li or $^{40}$K-$^{40}$K, with different trap oscillation frequencies, the phase separation discussed in this paper can be checked with the density measurement by absorption imaging the sample of ultracold atoms at tuning the repulsive interaction strength by Feshbach resonance. For example, for the case of spin-unpolarized $^6$Li system of $40$ particles with axial trap oscillation frequencies for the spin-down and spin-up species of $2\pi\times 10$ Hz and $2\pi\times 30$ Hz, respectively, the PS point will appear around $\lambda\approx 8$.

\section{Acknowledgements}
We thank A-Hai Chen for his help in producing the LSDA data. The hospitality
of the Condensed Matter Section of the Abdus Salam International Center for Theoretical Physics is acknowledged, where a part of the work was done.
This work was supported by the Natural Science Foundation of China under Grant No. 11174253 and the Zhejiang Provincial Natural Science Foundation under Grant No. R6110175.

\end{document}